\documentclass[default]{sn-jnl}

\usepackage{graphicx}%
\usepackage{multirow}%
\usepackage{amsmath,amssymb,amsfonts}%
\usepackage{amsthm}%
\usepackage{mathrsfs}%
\usepackage[title]{appendix}%
\usepackage{xcolor}%
\usepackage{textcomp}%
\usepackage{manyfoot}%
\usepackage{booktabs}%
\usepackage{algorithm}%
\usepackage{algorithmicx}%
\usepackage{algpseudocode}%
\usepackage{listings}%
\usepackage{orcidlink}%
\usepackage{mdframed}%
\usepackage{glossaries}%
\usepackage{todonotes}%
\usepackage{natbib}
\usepackage{subcaption}
\usepackage{array}
\usepackage{tabularx}
\usepackage{makecell}
\usepackage{geometry}
\usepackage{pgfplots}
\usepackage{pdflscape}
\usepackage{setspace}

\newmdenv[
  	backgroundcolor = black!9,
  	linecolor = black!30,
 	linewidth = 2.5pt,
 	leftmargin = 0,
 	rightmargin = 0,
 	innerleftmargin = 7.5pt,
 	innerrightmargin = 7.5pt, 
 	innertopmargin = 7.5pt,
 	innerbottommargin = 7.5pt,
 	topline=false,
 	bottomline=false,
 	rightline=false,
 	leftline=true
]{DefinitionFrameEnvironment}

\newcounter{DefinitionFrameCounter}

\makeglossaries

\newacronym{AI}{AI}{artificial intelligence}
\newacronym{ML}{ML}{machine learning}
\newacronym{LLM}{LLM}{large language model}
\newacronym{EHR}{EHR}{electronic health record}
\newacronym{HCI}{HCI}{human–computer interaction}
\newacronym{API}{API}{application programming interface}
\newacronym{FHIR}{FHIR}{Fast Healthcare Interoperability Resources}
\newacronym{LLMonFHIR}{\gls{LLM} on \gls{FHIR}}{LLM on FHIR}
\newacronym{CRediT}{CRediT}{Contributor Roles Taxonomy}

\glsunset{LLMonFHIR}

\glsdisablehyper

\begin{document}

\title[LLM on FHIR - Demystifying Health Records]{\vspace{-1.5cm}LLM on FHIR - Demystifying Health Records}

\author*[1]{\fnm{Paul} \sur{Schmiedmayer}\orcidlink{0000-0002-8607-9148}}\email{schmiedmayer@stanford.edu}
\author[1]{\fnm{Adrit} \sur{Rao}\orcidlink{0000-0002-0780-033X}}
\author[1]{\fnm{Philipp} \sur{Zagar}\orcidlink{0009-0001-5934-2078}}
\author[1]{\fnm{Vishnu} \sur{Ravi}\orcidlink{0000-0003-0359-1275}}
\author[1, 2]{\fnm{Aydin} \sur{Zahedivash}\orcidlink{0000-0001-6835-1139}}
\author[3]{\fnm{Arash} \sur{Fereydooni}\orcidlink{0000-0002-2519-5137}}
\author[1, 3]{\fnm{Oliver} \sur{Aalami}\orcidlink{0000-0002-7799-2429}}
\affil[1]{\orgdiv{Byers Center for Biodesign}, \orgname{Stanford University}, \orgaddress{\street{318 Campus Drive}, \city{Stanford}, \postcode{94305}, \state{CA}, \country{USA}}}
\affil[2]{\orgdiv{Department of Pediatrics}, \orgname{Stanford University}, \orgaddress{\street{453 Quarry Road}, \city{Stanford}, \postcode{94305}, \state{CA}, \country{USA}}}
\affil[3]{\orgdiv{Division of Vascular and Endovascular Surgery, Department of Surgery}, \orgname{Stanford University}, \orgaddress{\street{780 Welch Road}, \city{Stanford}, \postcode{94305}, \state{CA}, \country{USA}}}

\maketitle
\thispagestyle{empty}

\begin{abstract}

\textbf{Objective:} To enhance health literacy and accessibility of health information for a diverse patient population by developing a patient-centered \gls{AI} solution using \glspl{LLM} and \gls{FHIR} \glspl{API}.\\
\textbf{Materials and Methods:} The research involved developing \gls{LLMonFHIR}, an open-source mobile application allowing users to interact with their health records using \glspl{LLM}. The app is built on Stanford's Spezi ecosystem and uses OpenAI's GPT-4. A pilot study was conducted with the SyntheticMass patient dataset and evaluated by medical experts to assess the app's effectiveness in increasing health literacy. The evaluation focused on the accuracy, relevance, and understandability of the \gls{LLM}'s responses to common patient questions.\\
\textbf{Results:} LLM on FHIR demonstrated varying but generally high degrees of accuracy and relevance in providing understandable health information to patients. The app effectively translated medical data into patient-friendly language and was able to adapt its responses to different patient profiles. However, challenges included variability in LLM responses and the need for precise filtering of health data.\\
\textbf{Discussion and Conclusion:} \glspl{LLM} offer significant potential in improving health literacy and making health records more accessible. \gls{LLMonFHIR}, as a pioneering application in this field, demonstrates the feasibility and challenges of integrating \glspl{LLM} into patient care. While promising, the implementation and pilot also highlight risks such as inconsistent responses and the importance of replicable output. Future directions include better resource identification mechanisms and executing \glspl{LLM} on-device to enhance privacy and reduce costs.

\end{abstract}

\clearpage
\pagenumbering{arabic} 
\glsresetall


\section{Background and Significance}
\label{sec:introduction}

The US Government's Healthy People 2030 initiative's overarching goal is to \textit{"eliminate health disparities, achieve health equity, and attain health literacy to improve the health and well-being of all."}~\cite{national2019criteria}
The initiative promotes health literacy as a key goal to improve health and well-being over the next decade~\cite{national2019criteria}. 
Initiatives like these underscore the profound impact enhanced health literacy could have on individuals' overall health and well-being over the next decade.
Health literacy, as defined by the Health Resources \& Services Administration, extends beyond mere access to health information; it encompasses the critical ability to find, comprehend, and effectively utilize health-related information~\cite{hrsa2013health}.

The introduction of HL7 Fast Healthcare Interoperability Resources (FHIR) APIs and the enactment of the 21st Century Cures Act mark significant milestones in this journey~\cite{us2016century}.
These patient-facing APIs enable individuals to have access to their electronic health records stored across care providers and have the potential to revolutionize how individuals interact with their electronic health records, laying the groundwork for a new patient empowerment and engagement era. 
Focusing on understanding and acting on these insights is the crucial next step to positively impact patients' health, improve health outcomes, reduce clinical workloads, and democratize access to health data~\cite{graham2008do}.
The emerging field of mobile health data further amplifies these opportunities and challenges, offering a rich source of real-time, personalized health insights while intensifying concerns over data management, privacy, and security~\cite{shaw2018beyond, johnson2020individuals}.
Nevertheless, despite all these advancements, initiatives to provide access to health records infused with verbose medical jargon and uninterpreted mobile health data have not yet led to a giant leap toward transforming patient health outcomes.

Given that limited English proficiency has been implicated in poor health outcomes in the US~\cite{sentell2012low}, fostering understandability and equal access to health data without hurdles such as language barriers, medical knowledge~\cite{bachmann2007medicalknowledge}, and limited access is essential to improve health equity and social impact.
Our project proposes making health literacy radically accessible, understandable, and actionable, particularly to under-resourced communities. 
The advent of \glspl{LLM}~\cite{vaswani2017attention} in healthcare holds tremendous potential to democratize health literacy by augmenting human knowledge and capabilities.
Large language models like OpenAI's GPTs~\cite{openai2023gpt4, brown2020language} can address these challenges and unlock improved comprehension and engagement with patient access to electronic health records and personal health data.
We hypothesize that \glspl{LLM} can significantly elevate health literacy by enhancing human knowledge and skills.
While significant research is conducted to support physician and electronic health record system integrations, we are taking an essential human-centered approach: By proposing patient-centered, privacy-preserving AI solutions, we aim to transform how health information is accessed and understood, catering to a diverse patient population.
This represents a pivotal shift in healthcare, with the power to reduce the strain on medical professionals and place patients at the helm of their health journey ~\cite{lorkowski2021shortage}.
We see this approach as a crucial cornerstone with immediate clinical impact, reducing workloads and enabling focus for medical professionals while empowering patients to be the steward of their health journey.

Various studies have explored the use of \glspl{LLM} in healthcare.
These include aiding clinicians with decision-making and documentation \cite{shah2023creation}, and enhancing medical education \cite{clusmann2023future}.
Additionally, \glspl{LLM} show promise in patient-facing applications, particularly in answering medical questions.
Preliminary validations have assessed the efficacy of  \glspl{LLM} such as GPT in this context \cite{johnson2023assessing, nori2023capabilities}.
Integrating these models into the mobile health ecosystem from a digital health perspective can have the potential to improve medical literacy and understanding of one's health.
Patients often struggle to fully understand their health information due to the vast amount of data, compounded by language barriers and widespread lack of medical literacy \cite{wittink2018patient, hickey2018lowhealthliteracy}.
Incorporating \glspl{LLM} tailored to one's personal \gls{FHIR} records, within a mobile app, has the potential to increase the accessibility and understanding of one's health.
Our software, \gls{LLMonFHIR}~\cite{Schmiedmayer_Stanford_LLM_on}, seeks to develop solutions that address these issues, paving the way for responsible, secure, and effective use of \glspl{LLM} and mobile health data in healthcare as a critical step in enabling the next generation of digital health advancements.
We demonstrate the development and considerations of this patient-facing \gls{LLM}-based tool and highlight its applicability to make health records available to patients while evaluating its efficacy and potential challenges using sophisticated expert reviews.
Our pilot study focuses on an automatically synthesized user cohort of simulated \gls{FHIR} resources for patients with cardiovascular diseases and expert reviews of common questions asked by patients.

\section{Materials and Methods}
\label{sec:materialsandmethods}

The software developed as part of this research project is an open-source application licensed under the MIT license\footnote{\url{https://github.com/StanfordBDHG/LLMonFHIR/blob/main/LICENSE.md}} and builds on the foundational open-source \textit{Stanford Spezi} ecosystem~\cite{schmiedmayer2023spezi} and GPT-4~\cite{openai2023gpt4} in its \texttt{gpt-4-1106-preview}\footnote{OpenAI GPT-4 Turbo: \url{https://openai.com/blog/new-models-and-developer-products-announced-at-devday}} version.
The evaluation in a pilot study of the \gls{LLM}-based solution relies on the SyntheticMass patient and population health dataset generated by the Synthea\texttrademark~generator~\cite{walonoski2017synthea}.

\subsection{\gls{LLMonFHIR}}
\label{subsect:llmonfhir}

\gls{LLMonFHIR} is an open-source Swift-based iOS application that demonstrates the ability to interactively converse with one's FHIR records using \glspl{LLM}~\cite{Schmiedmayer_Stanford_LLM_on}.
The \gls{LLMonFHIR} app enables users to ask questions to an \gls{LLM}, which then generates responses based on the user's personal health records retrieved from an \gls{EHR}.
The app uses the open-source Stanford Spezi ecosystem~\cite{schmiedmayer2023spezi}, the successor of CardinalKit~\cite{aalami2023cardinalkit}, allowing the modular component-based development of digital health applications.
The Stanford Spezi~\cite{schmiedmayer2023spezi} ecosystem is a collection of software subsystems formulating modules that enable reusability and high-quality mobile computing applications, specifically targeting the digital health research and translation ecosystem.

\begin{figure}[h]
    \centering
    \begin{subfigure}[b]{0.3\textwidth}
        \centering
        \includegraphics[width=\textwidth]{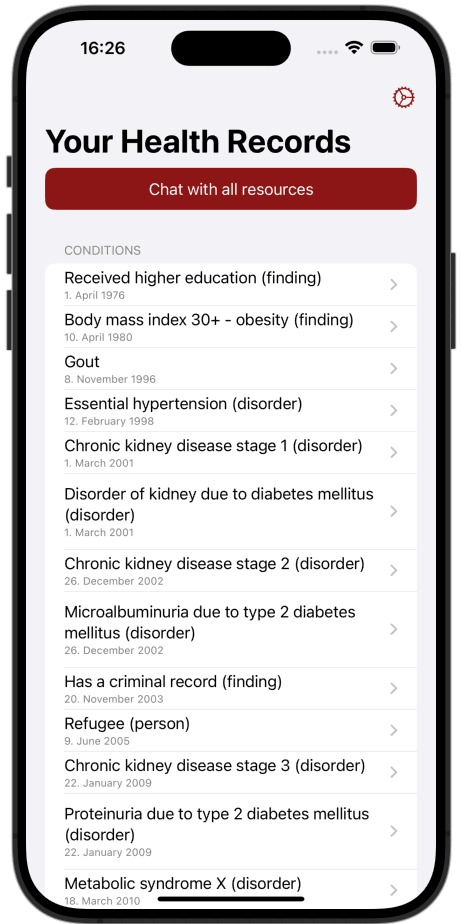}
        \caption{Overview of all available FHIR records}
        \label{fig:llm_fhir_overview}
    \end{subfigure}
    \hfill
    \begin{subfigure}[b]{0.3\textwidth}
        \centering
        \includegraphics[width=\textwidth]{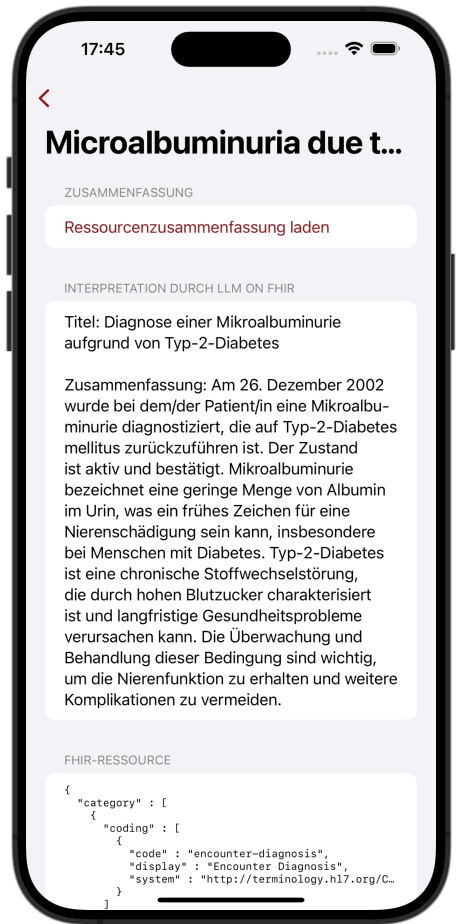}
        \caption{Interpretation and Summary of a single resource}
        \label{fig:llm_fhir_interpretation}
    \end{subfigure}
    \hfill
    \begin{subfigure}[b]{0.3\textwidth}
        \centering
        \includegraphics[width=\textwidth]{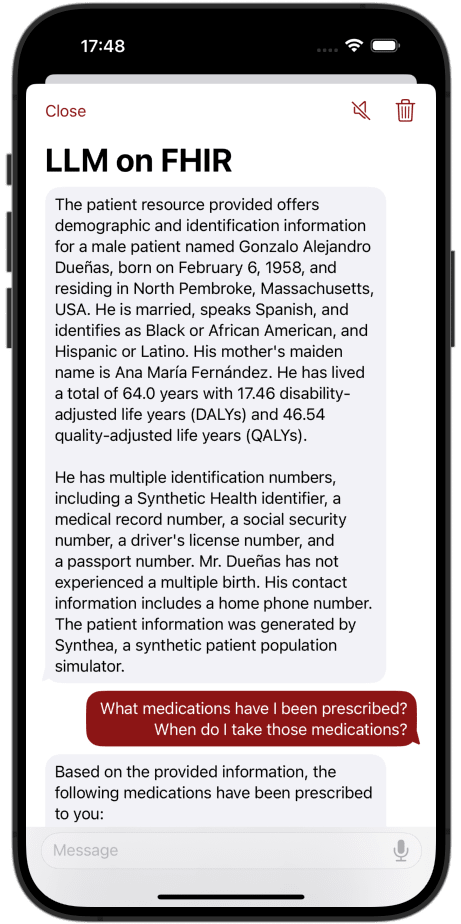}
        \caption{Interactive Chat with all FHIR health records}
        \label{fig:llm_fhir_chat}
    \end{subfigure}
    \caption{Screenshots of the \gls{LLMonFHIR} iOS Application}
    \label{fig:llm_fhir_screenshots}
\end{figure}

\subsubsection{FHIR Resources Data Flow}
\label{subsubsect:llmonfhir}

Enabled by \textit{SpeziHealthKit}~\cite{schmiedmayer2023spezihealthkit}, \gls{LLMonFHIR} connects with the Apple Health app using HealthKit\footnote{Apple HealthKit: \url{https://developer.apple.com/documentation/healthkit}} to obtain users' \gls{FHIR}-encoded health records across various hospitals and institutions.
The Apple Health app manages authentication and data fetching, simplifying the development of patient-facing solutions accessing health records retrieved from the \gls{EHR}.
In addition, \gls{LLMonFHIR} provides a mechanism to load patient resources from a \gls{FHIR} bundle used for testing and validating the application's functionality. 

Once health records are delivered to the application, \gls{LLMonFHIR} enables users to interact with individual resources of their interest and to inspect the underlying FHIR data in detail (\autoref{fig:llm_fhir_overview}).
Users can tap on any specific health record to navigate to a view providing a summary and interpretation in a particular record using \gls{LLM}s.
The GPT-4 interaction was prompt-engineered to provide short, concise summaries and longer-form interpretations based on the FHIR data in JSON format for the single resource (shown in \autoref{fig:llm_fhir_interpretation}) to help users understand the essence of the records.
The \textit{SpeziLLM} module~\cite{schmiedmayer2023spezillm} allows developers to interface with \gls{LLM}s in mobile applications ranging from cloud services to the execution of smaller models in a local execution environment.

\subsubsection{All \gls{FHIR} Resources Chat}
\label{subsubsect:allfhirresourceschat}

The key health literacy enabler and element validated in \autoref{sec:casesresultstudy} is a general interface allowing users to interact with their health records using an LLM-based chat interface (as shown in \autoref{fig:llm_fhir_chat}).
To further address the challenges posed by health literacy and improve the accessibility of \gls{LLMonFHIR}, voice and translation capabilities have been integrated into the application.
Voice-based interaction is facilitated across all chat interfaces within \gls{LLMonFHIR}, using the \textit{SpeziSpeech} module~\cite{schmiedmayer2023speech} from the Spezi ecosystem.
To allow for the use of different languages in \gls{LLMonFHIR}, we ensure that the \gls{LLM} has access to the user's locale and can respond to questions in the user's preferred language.
We enable full app translations in English, Spanish, Chinese, German, and French that translate all titles, descriptions, and the onboarding flow.

The \gls{LLM}-based interactions are facilitated using a \gls{LLM} function-calling mechanism, allowing the \gls{LLM} to request and retrieve information about specific health records by choosing from a list of predefined identities.
The identifiers are made up of a triplet of information detailed in the \gls{LLM} function call description and feature the (1) resource type, e.g., a medication request or observation, (2) display name of the resource computed based on the resource type, e.g., the title of the medication for a medication request, and (3) date that best describes the \gls{FHIR} resource such as the date of an observation or start date of a medication request.
The \gls{LLM} system prompt starting the conversation and introducing the function call mechanism can be found in \autoref{fig:llminterpretationprompt}.

\begin{figure}[h]
    \centering
\begin{lstlisting}
You are the LLM on FHIR application. Your task is to interpret FHIR resources from the user's clinical records.

Throughout the conversation with the user, use the "get_resources" function to obtain the FHIR health resources necessary to answer the user's question properly. For example, if the user asks about their allergies, you must use the "get_resources" function to output the FHIR resource titles for allergy records so you can then use them to answer the question. The end goal is to answer the user's question in the best way possible while taking the FHIR resources obtained using "get_resources" into consideration. Leave out any technical details like JSON, FHIR resources, and other implementation details of the underlying data resources.
Only request relevant resources and focus on recent resources. Try to reduce the number of requested resources to a reasonable scope.
E.g., if the user asks about their Medication, it would be recommended to request all MedicationRequest FHIR resource types.

Interpret the resources by explaining the data relevant to the user's health. Explain the relevant medical context in a language understandable by a user who is not a medical professional, ideally at a 5th-grade reading level. You should provide factual and precise information in a compact summary in short responses.

The patient resource is provided as the initial context. Ensure that you leave out anything like SSN, passport number, and telephone number. 
Do not introduce yourself at the beginning, and immediately return a summary of the user based on the FHIR patient resources. Do not request any FHIR resources using the "get_resources" function for your first message. The initial summary should be short and simple; stay at a high level. End with a question asking the user if they have any questions. Make sure your response is in the same language the user writes to you in, not the one stated in the patient resource. The tense should be present.
\end{lstlisting}
    \caption{Main multiple resources prompt provided as default in the \gls{LLMonFHIR} application. The prompt instructs the model to use the provided LLM function calling mechanisms and establishes a base expectation of the interaction with the user.}
    \label{fig:llminterpretationprompt}
\end{figure}

Health resources are pre-filtered to avoid exceeding the \gls{LLM} context limit and to reduce the number of choices for the model to a reasonable size.
For medications, only active and outpatient medications are included in the function call considerations.
Resources like observations, lab values, and conditions are filtered to reduce them to the most recent occurrence of each type of resource, focusing on the patient's need to identify and explain recent occurrences and procedures.
The patient record providing detailed information about a patient is injected at the beginning of each \gls{LLM} interaction, ensuring that the model understands the rough patient context adequately.
The model can use the function calling mechanism at any time to request information about a specific resource or a type of resource.
\gls{FHIR} resources are subsequently queried and processed using the summarization mechanism (as described in \autoref{subsubsect:llmonfhir}) to create a representative text of fewer than 100 words that are then fed into the larger \gls{LLM} chat context forming the interaction with the patient.

\subsection{Methods}
\label{subsect:methids}

To evaluate the ability of \gls{LLMonFHIR} to increase health literacy in a pilot study, we asked a group of four Medical Doctors at the Stanford School of Medicine to judge the responses from the \gls{LLM} in the interactive chat mode with all available health records (see \autoref{subsubsect:allfhirresourceschat}).
We selected a cohort of six exemplified FHIR patient data sets from the \textit{SyntheticMass dataset (Version 2)} generated by \textit{Synthea}~\cite{walonoski2017synthea} to allow for a reproducible and unbiased evaluation of the responses from the Language Model.
SyntheticMass has been used for many preliminary machine learning model evaluations \cite{chen2022simulation, walonoski2020synthea, rankin2020reliability}.
To select representative patients, the entire dataset was partitioned into ten individual buckets consisting of a unique cardiovascular condition/procedure, given that cardiovascular conditions were the specialty of most of our expert reviewers.
After partitioning the entire dataset and excluding patients that did not fit the criteria, a balanced set of patients was selected by ensuring that all patients were still alive, at least two patients had allergies, and that the selected patients resulted in a representative balance of self-reported gender, ethnic background, and age groups ranging from young patients (8 years old) to elderly patients (the most senior patient is 82 years old).
The list of the selected patient data set and a summary of their corresponding conditions, allergies, and medications can be found in \autoref{tab:selected_synthetic_patients}.

\begin{table}[]
\renewcommand{\arraystretch}{1.5}
\centering
\resizebox{\textwidth}{!}{%
\begin{tabular}{|m{1.45cm}|c|c|c|c|c|}
\hline
\textbf{Name} & \textbf{Sex} & \textbf{Age} & \textbf{Conditions} & \textbf{Allergies} & \textbf{Medications} \\ \hline
Beatris270 Bogan287 & F & 8 years & \makecell{Aortic valve stenosis (disorder) \\ Perennial allergic rhinitis \\ Atopic dermatitis} & \makecell{Latex (substance) \\ Bee venom (substance) \\ Mold (organism) \\ House dust mite (organism) \\ Animal dander (substance) \\ Grass pollen (substance) \\ Tree pollen (substance) \\ Aspirin} & \makecell{Fexofenadine hydrochloride 30 MG Oral Tablet \\ Epinephrine 1 MG/ML Auto-Injector 0.3 ML} \\ \hline

Milton509 Ortiz186 & M & 26 years & \makecell{Hypertension \\ Hypoxemia (disorder) \\ Stress (finding)} & - & \makecell{MedicationRequest \\ amLODIPine 2.5 MG Oral Tablet} \\ \hline

Edythe31 McDermott739 & F & 49 years & \makecell{Body mass index 30+ - obesity (finding) \\ Received higher education (finding) \\ Prediabetes \\ Anemia (disorder) \\ Victim of intimate partner abuse (finding) \\ Cardiac Arrest \\ History of cardiac arrest (situation)} & - & Jolivette 28 Day Pack \\ \hline

Gonzalo160 Dueñas839 & M & 65 years & \makecell{Body mass index 30+ - obesity (finding) \\ Gout
Essential hypertension (disorder) \\ Disorder of kidney due to diabetes mellitus (disorder) \\ Microalbuminuria due to type 2 diabetes mellitus (disorder) \\ Proteinuria due to type 2 diabetes mellitus (disorder) \\ Metabolic syndrome X (disorder) \\ Prediabetes \\ Anemia (disorder) \\ Ischemic heart disease (disorder) \\ Abnormal findings diagnostic imaging- \\ heart+coronary circulat (finding) \\ History of renal transplant (situation) \\ Medication review due (situation)} & - & \makecell{Simvastatin 20 MG Oral Tablet \\ Vitamin B12 5 MG/ML Injectable Solution \\ Clopidogrel 75 MG Oral Tablet \\ Hydrochlorothiazide 25 MG Oral Tablet \\ amLODIPine 2.5 MG Oral Tablet \\ Metoprolol succinate 100 MG 24 HR Extended Release Oral Tablet \\ Insulin isophane, human 70 UNT/ML / insulin, regular, \\ human 30 UNT/ML Injectable Suspension [Humulin] \\ Nitroglycerin 0.4 MG/ACTUAT Mucosal Spray \\ Tacrolimus 1 MG 24 HR Extended Release Oral Tablet} \\ \hline

Jacklyn830 Veum823 & F & 72 years & \makecell{Essential hypertension (disorder) \\ Miscarriage in first trimester \\ Ischemic heart disease (disorder) \\ Chronic kidney disease stage 3 (disorder) \\ Proteinuria due to type 2 diabetes mellitus (disorder) \\ Social isolation (finding) \\ Sprain (morphologic abnormality)} & - & \makecell{Nitroglycerin 0.4 MG/ACTUAT Mucosal Spray \\ Simvastatin 20MG Oral Tablet \\ Clopidogrel 75 MG Oral Tablet \\ 24 HR metoprolol succinate 100 MG \\ Extended Release Oral Tablet \\ Acetaminophen 325 MG Oral Tablet \\ Hydrochlorothiazide 25 MG Oral Tablet} \\ \hline

Allen332 Ferry570 & M & 82 years & \makecell{Chronic sinusitis (disorder) \\ Hypertension \\ Served in armed forces (finding) \\ Received higher education (finding) \\ Body mass index 30+ - obesity (finding) \\ Prediabetes \\ Anemia (disorder) \\ Opioid abuse (disorder) \\ Atrial Fibrillation \\ Neoplasm of prostate \\ Carcinoma in situ of prostate (disorder) \\ Chronic intractable migraine without aura \\ Victim of intimate partner abuse (finding) \\ Stress (finding) \\ Alzheimer's disease (disorder))} & \makecell{Animal dander (substance) \\ Penicillin V \\ Peanut (substance)} & \makecell{Galantamine 4 MG Oral Tablet \\ Warfarin Sodium 5 MG Oral Tablet \\ doxycycline hyclate 100 MG \\ 1 ML DOCEtaxel 20 MG/ML Injection \\ 0.25 ML Leuprolide Acetate 30 MG/ML Prefilled Syringe \\ lisinopril 10 MG Oral Tablet \\ Verapamil \\ Hydrochloride 40 MG \\ Digoxin 0.125 MG Oral Tablet} \\ \hline

\end{tabular}%
}
\caption{
Summary of the six selected exemplified FHIR patient data sets from the \textit{SyntheticMass dataset} generated by \textit{Synthea}. We use the \textit{SyntheticMass dataset} (Version 2) that consists of one million synthetic FHIR patient medical records from a simulated population in Massachusetts, USA.
}
\label{tab:selected_synthetic_patients}
\end{table}

Each patient can be selected in the \gls{LLMonFHIR} application, which loads all their \gls{FHIR} resources in the context of the application using the \gls{FHIR} bundle loading functionality.
Each reviewer selected the \textit{all resources chat} functionality, clears out the current context, and subsequentially asks the typical patient questions or questions that were determined based on the available information exposed by \gls{FHIR} APIs (\autoref{table:validationquestion}).
In total, we have evaluated 168 \gls{LLM} responses (6 patients x 7 questions x 4 repetitions) equally distributed to all expert reviewers while considering and explicitly inspecting the variations between the responses. 
Some of the questions build on top of each other while investigating the capabilities of \glspl{LLM} in a patient-facing solution to foster health literacy ranked in increasing complexity and challenge for the \gls{LLM}.

\begin{table}[ht]
    \centering
    \renewcommand{\arraystretch}{1.2}
    \begin{tabular}{|m{0.5cm}|m{11.5cm}|}
        \hline
        \textbf{ID} & \textbf{Question} \\
        \hline
        Q1 & What are my current medications and how should I be taking them? \\
        \hline
        Q2 & What are the most common side effects for each medication I am taking? \\
        \hline
        Q3 & Am I allergic to any of my medications? \\
        \hline
        Q4 & Can you summarize my current medical conditions? \\
        \hline
        Q5 & What are the health behaviors I should be incorporating into my daily routine to help with my conditions? \\
        \hline
        Q6 & Can you summarize my current medical conditions in German? \\
        \hline
        Q7 & What are my recent laboratory values, what do they mean, and how can I improve them? \\
        \hline
    \end{tabular}
    \caption{Questions that are sequentially asked to the \gls{LLMonFHIR} application to validate the responses' accuracy, understandably, and relevance. }
    \label{table:validationquestion}
\end{table}

Each answer is scored on a Likert score between 1 (Strongly Disagree) and 5 (Strongly Agree), asking four expert medical doctors if they agree that the response was factual and accurate based on the content provided by the \gls{FHIR} resources that were manually inspected and used when evaluating the responses of the \gls{LLM}.
In addition to accuracy, the four experts also evaluated the responses in relevance to the question and information available to the model using the same scale.
The third aspect determined the understandability of the question for the target audience of a patient with no in-depth medical knowledge and their provided age.
To judge understandability, experts investigated the usage of medical terms that might not be known by patients and the quality of the instructions provided by the LLM for any suggestions or interpretations.
The answers to the translation question (Q6) were inspected by three German native speakers and translated using Google Translate to enable the medical experts to inspect the factual accuracy of the translated content.

An explicit limitation of the setup is that every expert interacts individually with the Language Model, resulting in slightly different answers from the model and somewhat deviating chat histories.
Each interaction with the \gls{LLM} was recorded using a screen recording. The tests were automated using Apple's user interface testing framework and are provided in the open-source implementation to enable full reproducibility of the results.
The automation interaction was designed to ensure that the LLM chat and context were cleared before asking the first question and that questions were always asked in the same sequential order (Q1 to Q7).

\section{Results}
\label{sec:casesresultstudy}

\begin{figure}[h]
\centering
\includegraphics[width=0.9\textwidth]{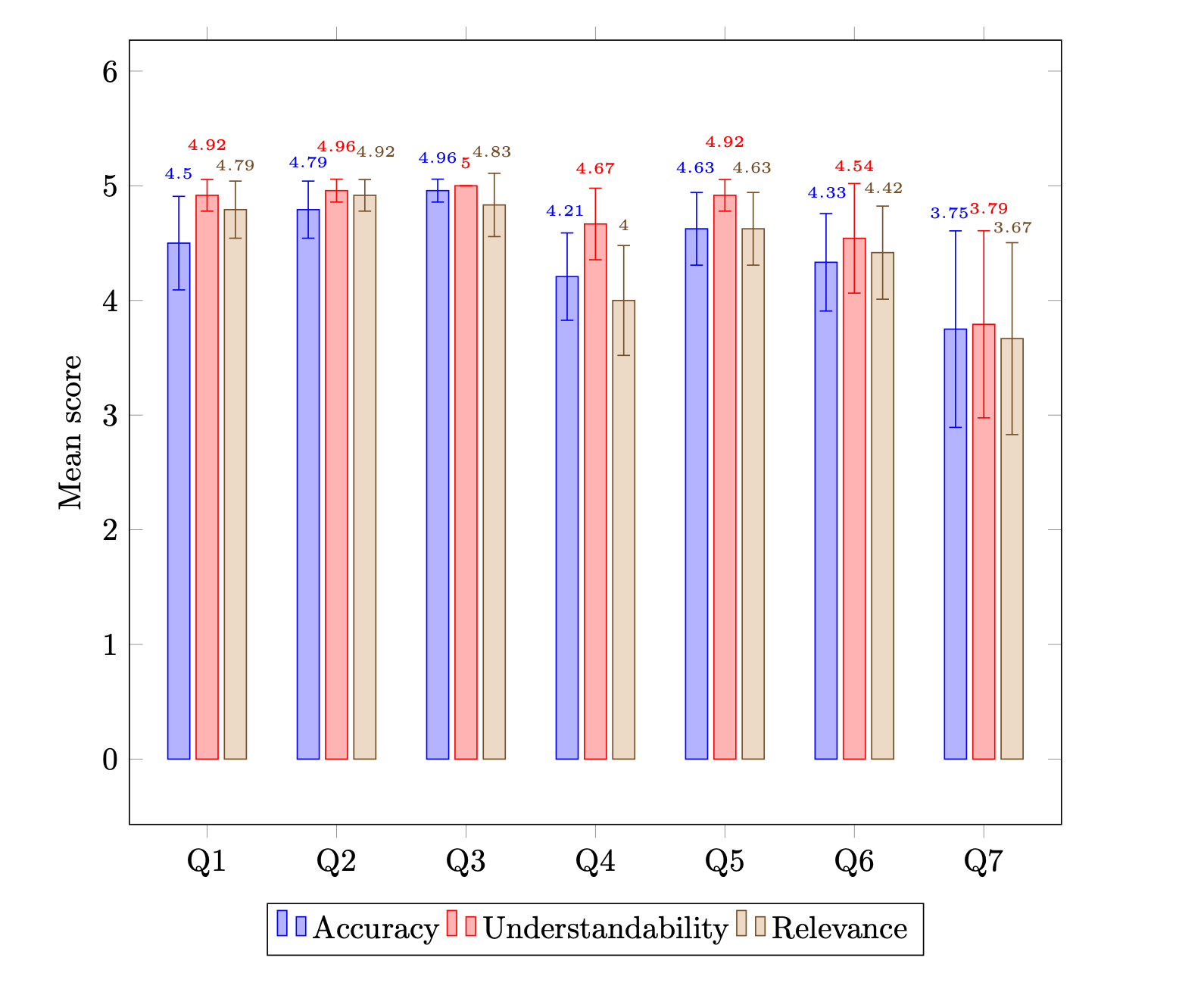}
\caption{Scoring of \gls{LLMonFHIR} answers by expert reviewers displaying the average response (added as an annotation) as well as the standard deviation for each question and subset of scores.}
\label{fig:rating_chart}
\end{figure}

The expert reviews demonstrated varying but generally high degrees of accuracy and relevance in providing understandable health information to patients.
The accuracy, understandably, and average relevance scores (see \autoref{subsect:methids}) and their standard deviations are displayed in \autoref{fig:rating_chart}.
The overall expert perception was that the app effectively translated medical data into patient-friendly language and was able to adapt its responses to different patient profiles.
However, challenges included variability in LLM responses and the need for precise filtering of health data.

Across the board, most experts scored the responses as accurate and especially understandable, highlighting the ability of the model to transform complex concepts into more concise and expressible patient-facing summaries.
Expert reviewers that deducted points in the scoring highlighted that the output of the model sometimes misses information included in the \gls{FHIR} resources and observed that the model did not request all necessary resources using the function-calling mechanism.

\begin{figure}[H]
    \centering
        \begin{subfigure}[b]{0.23\textwidth}
        \centering
        \includegraphics[width=\textwidth]{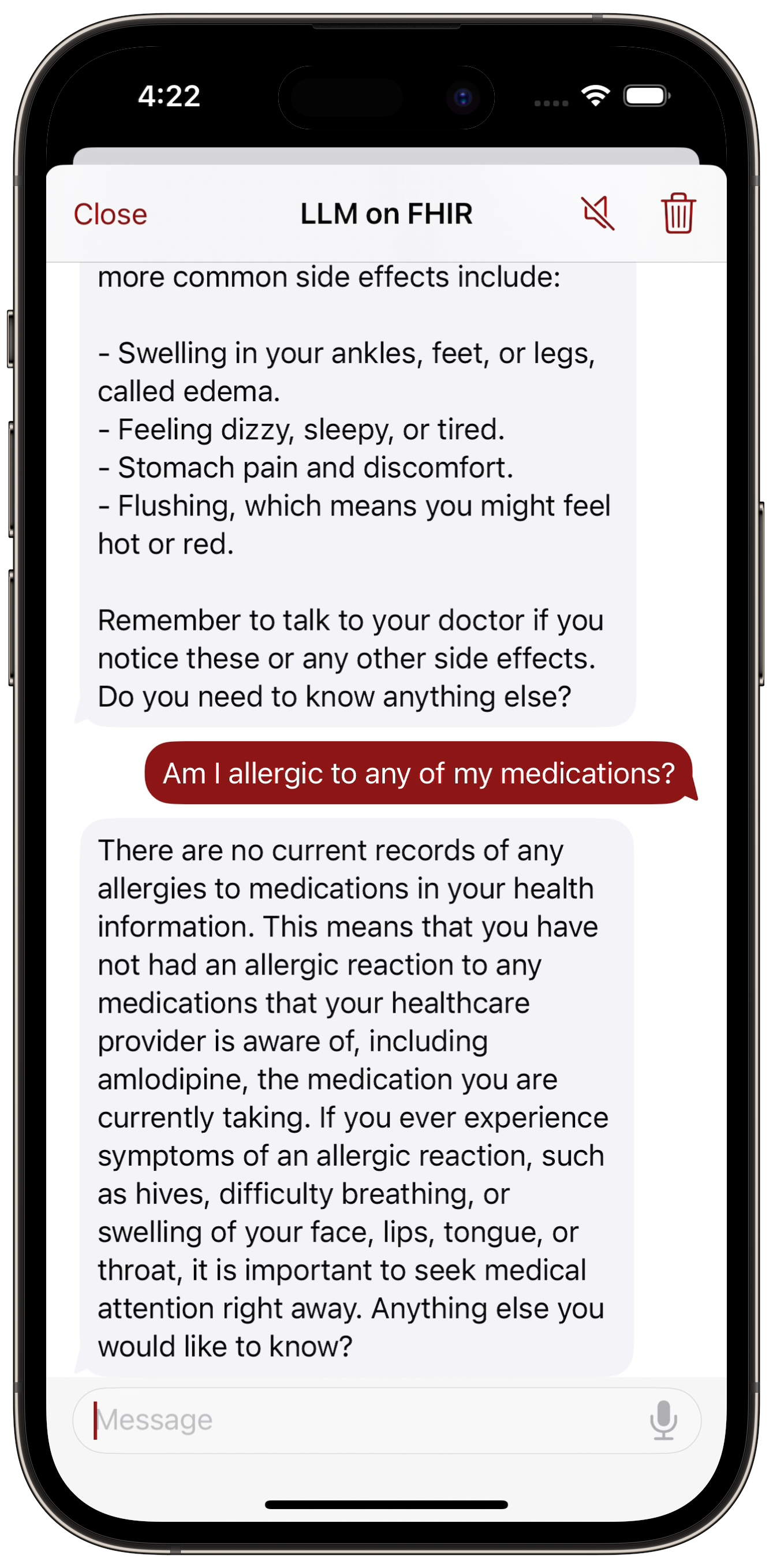}
        \caption{Answer to question 3, showing no hallucinations and correctly identifying no allergies.}
        \label{fig:llm_response_good_no_hallucination}
    \end{subfigure}
    \hfill
    \begin{subfigure}[b]{0.23\textwidth}
        \centering
        \includegraphics[width=\textwidth]{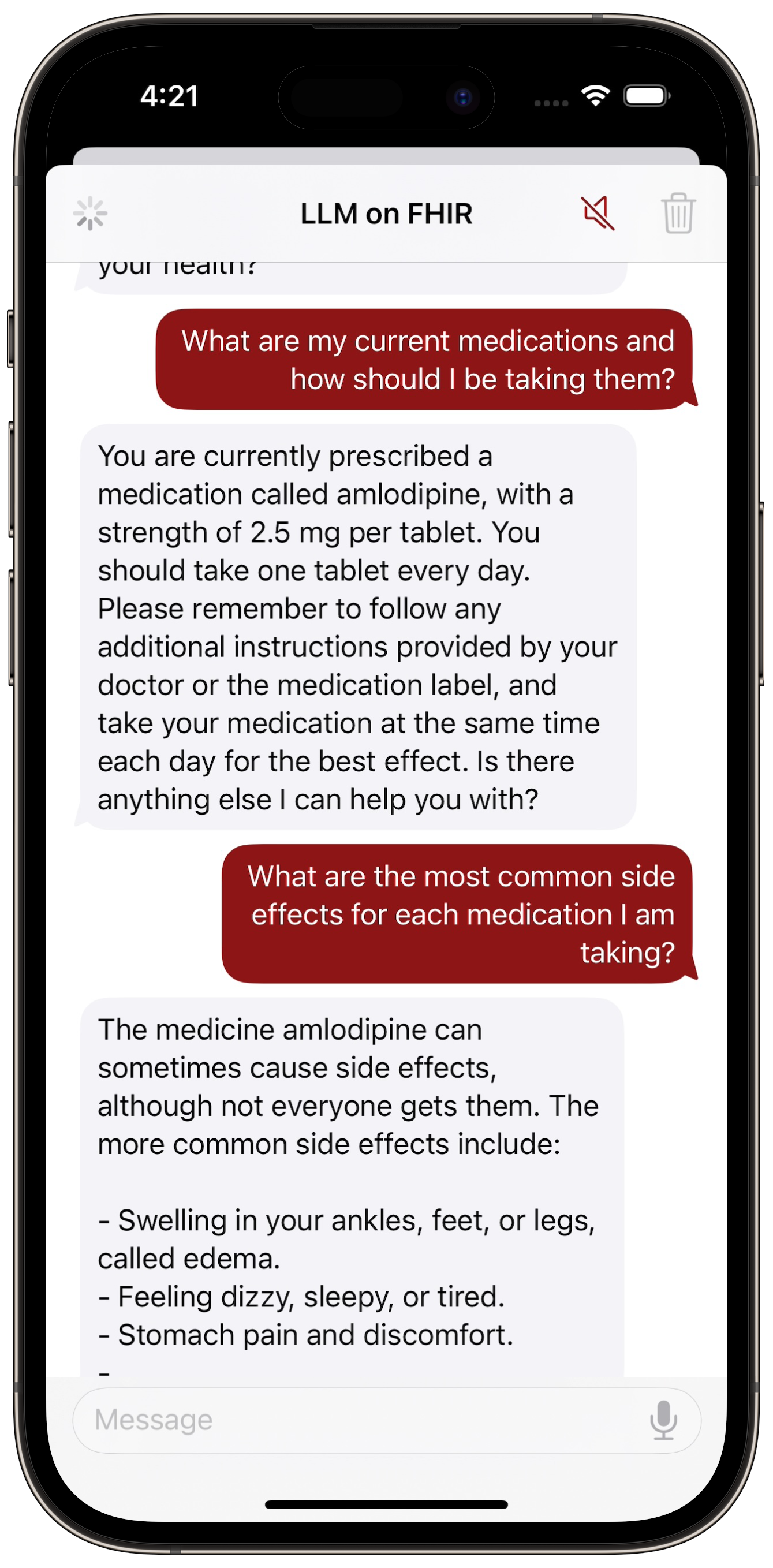}
        \caption{Answer to question 2, showing an accurate and relevant response listing all medications.}
        \label{fig:llm_response_good_accurate_relevant}
    \end{subfigure}
    \hfill
    \begin{subfigure}[b]{0.23\textwidth}
        \centering
        \includegraphics[width=\textwidth]{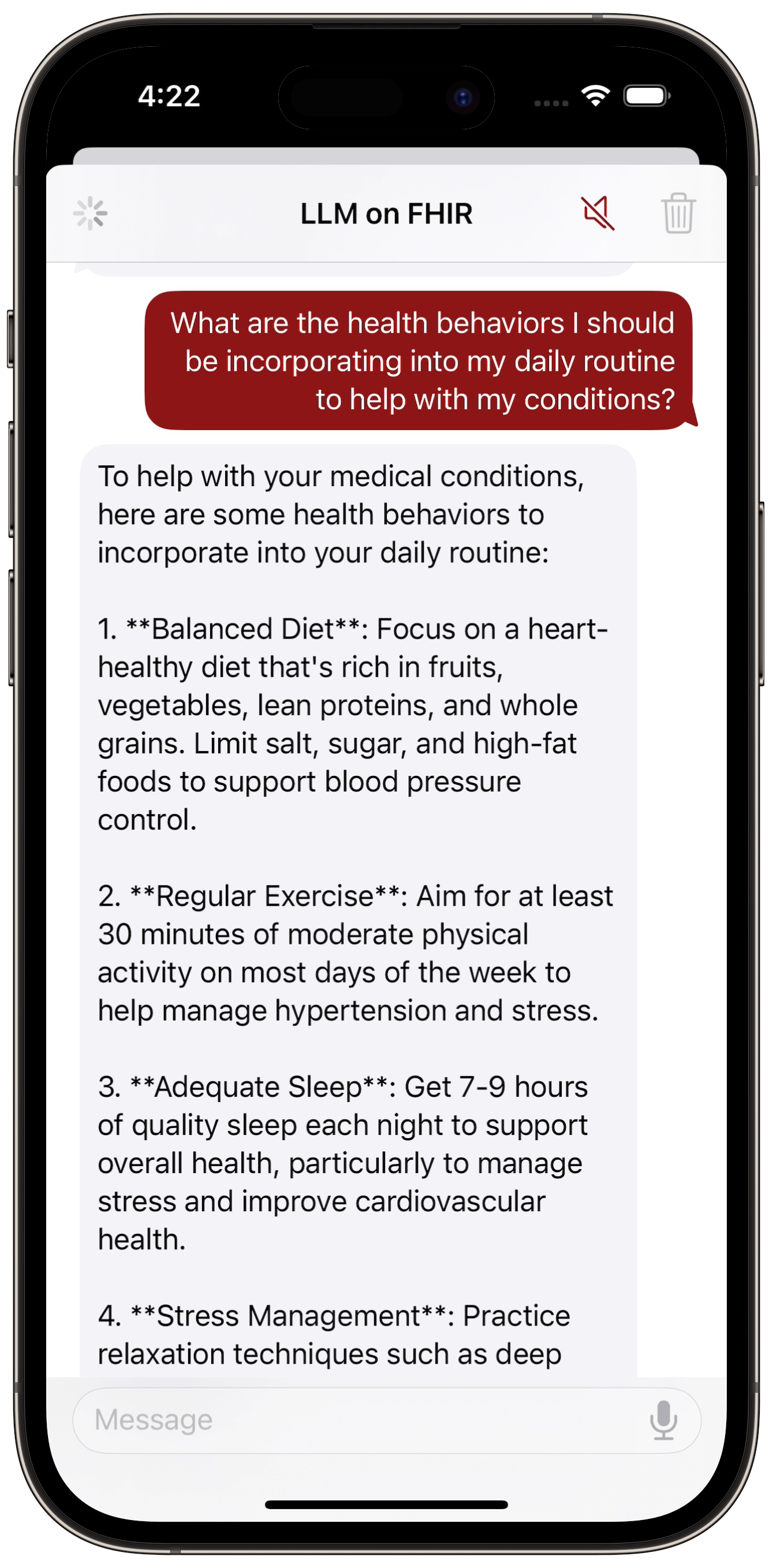}
        \caption{Answer to question 5, showing generic, verbose and irrelevant behavior suggestions.}
        \label{fig:llm_response_bad_irrelevance}
    \end{subfigure}
    \hfill
    \begin{subfigure}[b]{0.23\textwidth}
        \centering
        \includegraphics[width=\textwidth]{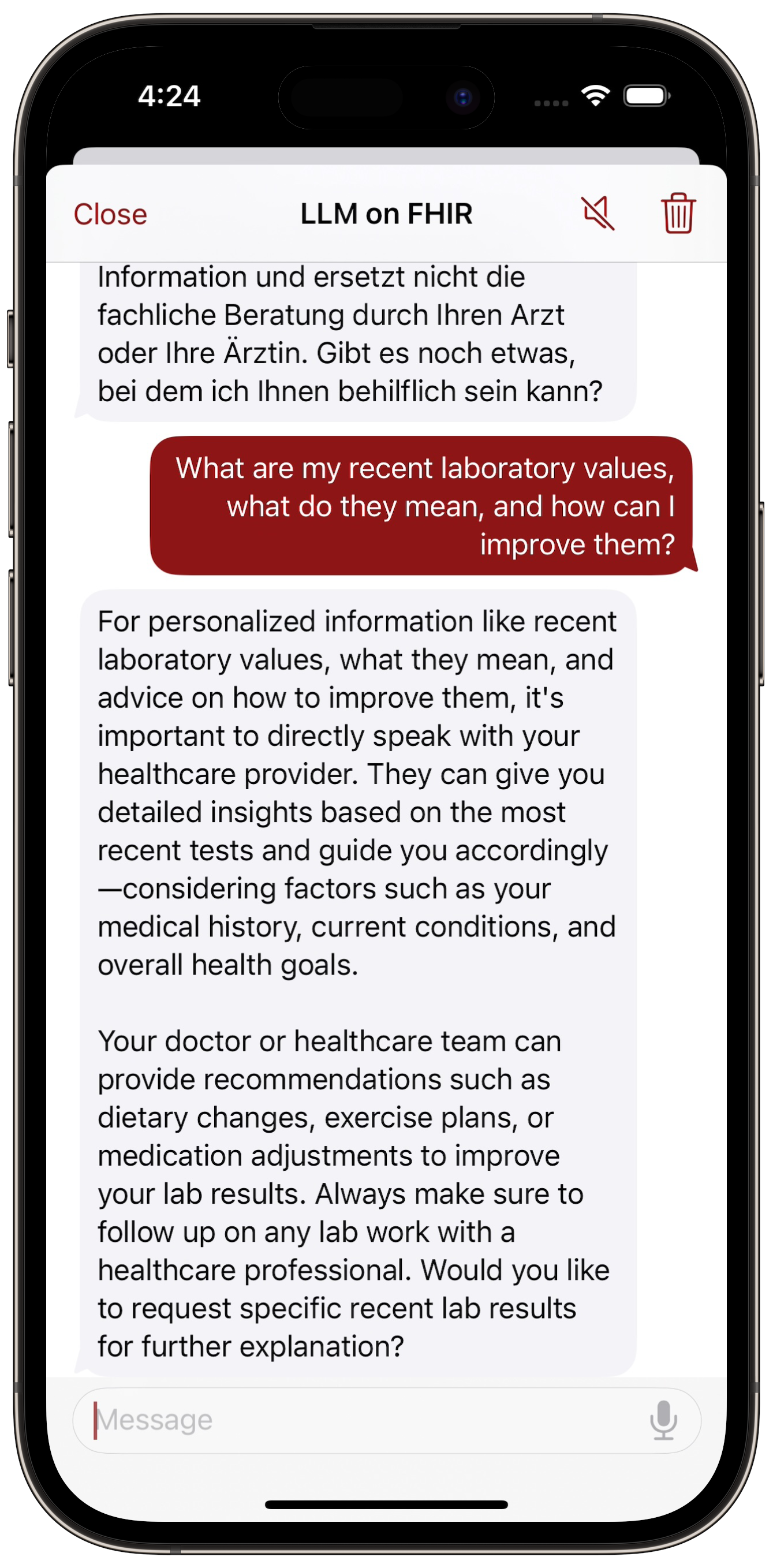}
        \caption{Answer to question 7, listing no relevant answer and requesting no \gls{FHIR} resources.}
        \label{fig:llm_response_bad_no_answer}
    \end{subfigure}
    \hfill
    \caption{Outputs of \gls{LLMonFHIR} judged by expert reviewers as part of the pilot study.}
    \label{fig:llm_responses}
\end{figure}

Questions 1, 2, and 3, focusing on medications, generally resulted in high ratings from the reviewers, highlighting the tone and understanding of the responses as well as correct context as demonstrated in \autoref{fig:llm_response_good_no_hallucination} and \autoref{fig:llm_response_good_accurate_relevant}.
The answers were seen as very clear and concise, providing a well-reasoned explanation for each medication, including its dosage and route for most medications.
Additionally, the \gls{LLM} demonstrated caution with certain medications, such as insulin, recommending adherence to prescribed instructions.
The \gls{LLM} always identified if the patient had no allergies and clearly reported this in the chat interface.
The main expert feedback stemmed from missing information about dosages and routes taken in some answers, one misassociation  where the model reported that insulin is used for prediabetes despite the patient having a type 2 diabetes diagnosis, and instances of rare side effects reported as common.

In investigating questions 4, 5, and 6, expert reviewers deducted points where the model provided out-of-context information, including conditions such as education status and refugee status which would traditionally fall under a social history as opposed to a medical history.
The \gls{LLM}'s understanding of recent conditions and resulting implications on current behavior were criticized by some experts as not relevant and intermixed in irrelevant information or too generic advice as seen in \autoref{fig:llm_response_bad_irrelevance}.
We performed a sub-analysis of Q4 (summarizing health conditions) to assess variability in answers provided by the model to the same question asked by the four different reviewers. 
Never were two responses exactly the same. 
Omission of conditions was noted in an estimated 20\% of the responses.
Patterns of omission were not consistent. 
No obvious or significant hallucinations were noted. 

Question 6, focusing on translating information into another language, was generally perceived as correct and understandable for native German speakers while being close to the English answer to previous questions and lacking the typical nuances of interaction we perceived in English, e.g., adjusting the responses to the age and background of the patient (see \autoref{sec:discussion}). 
In addition, in one instance, the model only translated condition names and lost the remainder of the context.

The most significant variation of \gls{LLM}-performance was perceived in question 7, mostly stemming from the challenges of the model to retrieve insightful and applicable lab results from the \gls{FHIR} resources using the function-calling mechanism as shown in \autoref{fig:llm_response_bad_no_answer}.
Due to the missing context, several answers have neither been rated as accurate nor understandable as most of them did not provide any insightful information.
When the \gls{LLM} correctly identified lab values, expert reviewers perceived most of the summaries as accurate and ranges for lab values and interpretations as appropriate and valuable insights, while actionable insights were not always provided.
Ranges provided by the model were mostly perceived as accurate, while interpretations of these values in correlation to the lab results were sometimes too strict, e.g., marking values as too high when they were only slightly out or range.
In one instance, the model was noted to avoid answering the question completely and referred the patient to their healthcare provider.
\section{Discussion}
\label{sec:discussion}

\gls{LLMonFHIR} presents a novel approach to address the challenges of health literacy and the accessibility of health data for patients.
The application showcases the potential to augment clinicians' correspondence with patients regarding straightforward and frequently asked questions utilizing patient-facing \gls{FHIR} \glspl{API} facilitated by the 21st Century Cures Act~\cite{us2016century}.
By leveraging foundational \glspl{LLM} like GPT-4 and mechanisms to feed external information into the \gls{LLM} context, such as function calling, \glspl{LLM} can be used to aggregate and transform information stored in the \gls{EHR} and present it in a more accessible manner.
The opportunities become especially apparent when observing \glspl{LLM} interacting and interfacing with a wide range of patients in different age ranges, socioeconomic backgrounds, and levels of health literacy who might otherwise find it challenging to understand their health data and records.
The pilot study outlined in \autoref{sec:casesresultstudy} demonstrates the app's ability to translate complex information and provide context in a personalized and patient-appropriate setting.
To our knowledge, this is the first application using \glspl{LLM} in a patient-facing application to explain and contextualize health records retrieved using \gls{FHIR} \glspl{API}.

At the same time, using \glspl{LLM} in this context presents considerable risks and challenges that need to be investigated and addressed when allowing patients to interface with their health records.
This includes the challenge of reproducible answers and behavior when generating content using \glspl{LLM}.
While hallucinations were not common in our evaluation, we observed a noticeable variation in requests for patient resources for the same questions, resulting in different responses, sometimes excluding insightful patient information.
We aim to build an ecosystem of open-source tools interfacing with different \gls{LLM} providers and services addressing patient health literacy and health data processing challenges.
For example, the infrastructure could be used to automatically fill out intake forms or enroll patients in clinical trials using patient information processed by \glspl{LLM}.

As outlined in \autoref{sec:casesresultstudy}, the \gls{LLM} has varying, but generally high, degrees of accuracy and relevance when answering common questions based on a wide variety of patient profiles representing a typical subset of patients with multiple comorbidities.
Observing the variability of the \gls{LLM} 's responses demonstrated the possibilities of a new mechanism of \gls{HCI} by introducing nuances and context awareness that were previously challenging to achieve and hard to replicate across various domains.
The \gls{LLM} references patients in the responses and adjusts itself to a patient, as demonstrated in conversations with the 8-year-old patient.
The model starts to use more straightforward and child-friendly language in explaining complex medical information, even referring to parents or other adults within the generated outputs.

A key learning from building, using, and validating the application of \glspl{LLM} in the health literacy context is the necessity to strictly filter and automatically preprocess any data fed to the \gls{LLM}.
This challenge arises due to the limited context size and the hard-to-predict reasoning of requesting resources using function calls.
We continuously iterated over the process while developing and testing the application with synthetic patient data.
We uncovered good mechanisms to calculate identifiers for patient resources to provide enough context to the model, such as the triplet of information described in \autoref{subsect:llmonfhir}.
In addition, we developed a mechanism to remove outdated medications and older irrelevant observations that have been replaced by more recent insights and \gls{FHIR} resources.
As noted in the results, we still observed that irrelevant observations and conditions, such as social history facts, were surfaced by the model (e.g., criminal record, higher education, refugee status, need for medication review, and full-time employment).
These data points can be present in \gls{FHIR} record bundles but should not surface in a patient-facing context outlining the medical conditions.
In addition, the definition of \textit{recent} conditions was often perceived as inaccurate, highlighting the challenges of the \gls{LLM} of understanding and processing temporal correlations.

The perceived challenge of variability within the generated answers can be addressed by reducing the \gls{LLM} output temperature setting and adjusting other model parameters to fit better the individual context of the question in the patient correspondence.
In addition, initial seed definitions and other \gls{LLM} replicability efforts under active research are essential and become more relevant for medical applications.
Recent advancements by commercial offerings like GPT-4, such as reproducible outputs\footnote{\url{https://platform.openai.com/docs/guides/text-generation/reproducible-outputs}} will facilitate the applicability and replicability of these models in a patient context.
The eventual first adoption of these tools in a patient-facing context will require a manual review by clinicians.
We hypothesize it will gradually move to a more automated input and output risk assessment.
Additionally, we expect an expansion in the abilities of generative \gls{ML} models, enabling them to handle medical correspondences independently without needing extra supervision.

\section{Conclusion}
\label{sec:conclusion}

\glspl{LLM} provide tremendous opportunities for transforming, summarizing, making health records more accessible, and improving health literacy.
The \gls{LLMonFHIR} application demonstrates this promise while highlighting the current shortcomings and next steps in integrating \glspl{LLM} in clinical workflows and fitting the more remote patient care paradigm observed in the recent decade.
Automating typical patient interactions while safeguarding them from misinformation, hallucinations, and irrelevant information is the pathway to better, more scalable patient care.
By providing the application as an open-source project and sharing our results, we aim to ensure that other researchers can collaborate on a set of best practices, engineered prompts, and function-calling mechanisms establishing \gls{LLM} interactions with \gls{FHIR} resources.
The core infrastructure of \gls{LLMonFHIR} is provided as a reusable component using SpeziLLM and SpeziFHIR, enabling developers to utilize it to prototype and validate their research results.

We intend to keep enhancing \gls{LLMonFHIR} to democratize health data understandability and accessibility.
Nevertheless, given the sensitive nature of health data, we aim to shift the LLM execution environment from opaque, proprietary, and centralized cloud providers like OpenAI closer to the patient's device.
We hypothesize that executing open-source LLMs~\cite{touvron2023llama} in more trusted environments, like the patient's edge device, will help to mitigate privacy, trust, and financial challenges posed by cloud-based LLMs~\cite{yeung2023aichatbots, yuan2023llm, weidinger2021ethical}.
As an example, summarizations in \gls{LLMonFHIR}, as described in \autoref{subsect:llmonfhir}, can be performed by small, on-device models like Llama 2, which then serve as the input for more challenging interpretation tasks performed in the cloud-layer by more extensive and capable models.

Committing to the open-source and trust-building mechanism of collaborative software development, we plan to extend the \gls{LLM} capabilities in the Stanford Spezi ecosystem and facilitate the sharing of best practices and reusability in accessing \gls{EHR} data using \gls{FHIR} \glspl{API} to enable the scalability and usage of patient-empowering digital solutions in healthcare.

\backmatter

\section*{Acknowledgments}

We thank the Stanford Byers Center for Biodesign for supporting this project and our digital health research.

\section*{Author Contributions}
\gls{CRediT} author statement: \textbf{Paul Schmiedmayer}: Conceptualization, Software, Writing - (Original Draft, Review \& Editing), Supervision;
\textbf{Adrit Rao}: Conceptualization, Software, Writing - (Original Draft, Review \& Editing);
\textbf{Philipp Zagar}: Software, Writing - (Original Draft, Review \& Editing);
\textbf{Vishnu Ravi}: Software, Validation, Writing - Review \& Editing;
\textbf{Aydin Zahedivash}: Validation, Writing - Review \& Editing;
\textbf{Arash Fereydooni}: Validation, Writing - Review \& Editing;
\textbf{Oliver Aalami}: Conceptualization, Validation, Writing - (Original Draft, Review \& Editing), Supervision.

\section*{Declarations}

None of the authors have a competing interest. The \gls{LLMonFHIR} application and the Stanford Spezi ecosystem are open-source and licensed using the MIT license. Stanford Spezi is being used to foster a digital health ecosystem and teach the next generation of digital health leaders.

\section*{Data Availability}

We have included links to all the open-source software used in this manuscript. Further information and an overview of our open-source tools are available at \url{https://github.com/StanfordBDHG} and \url{https://github.com/StanfordSpezi}.

\pagebreak
\bibliographystyle{plainnat}
\bibliography{Bibliography}

\end{document}